\documentstyle[preprint,aps,epsf]{revtex}
\tightenlines
\begin{document}
\draft
\title{Fractional Populations in Sex-linked Inheritance}

\author{Seung Pyo Lee$^a$, Myung-Hoon Chung$^b$, Chul Koo Kim$^a$, and
       Kyun Nahm$^c$}

\address{$^a$Department of Physics and Institute for
         Mathematical Sciences, Yonsei University, Seoul 120-749, Korea}
\address{$^b$Department of Physics, Hong-Ik University,
         Chochiwon, Choongnam 339-800, Korea}
\address{$^c$Department of Physics, Yonsei University,
         Wonju 220-710, Korea}
\maketitle

\begin{abstract}
We study the fractional populations in chromosome inherited diseases. The
governing equations for the fractional populations are found and solved in the
presence of mutation and selection. The physical fixed points obtained are used
to discuss the cases of color blindness and hemophilia.
\end{abstract}
\pacs{87.10.+e, 42.62.Be}

\section{Introduction}
Many physical ideas are finding applications in complex biological systems
these days\cite{immu,mekjian}. Recently, we presented a theoretical scheme in
which one can investigate the ratios between the fractional population of blood
groups\cite{blood}. This method has some analogy with the physical concept of
renormalization and fixed points. In this paper, we extend the theory to
investigate the problem of sex-linked inheritance.

In sex-linked inheritance, there are five population groups, $XX$, $X^\prime X$,
$X^\prime X^\prime$, $XY$, and $X^\prime Y$, where $X$, $X^\prime$ and $Y$ represent
the normal female, the defective female, and the male chromosome, respectively.
The group of so called carrier female is characterized by $X^{\prime}X$. The
genetic rule is that sons receive $Y$ and daughters do $X$ or $X^\prime$ from
their fathers. Similarly, their mothers deliver $X$ or $X^\prime$ to sons and
daughters. It is also known that there exist mutations between $X$ and
$X^\prime$. Another important factor in the present problem is that the defective
groups, $X^\prime X^\prime$ and $X^\prime Y$, have disadvantages in surviving
and inheriting unlike in the case of the blood groups\cite{blood}.
This selection process should be taken
into account for any reasonable discussions. Therefore, we consider the
inheritance of sex-linked disease in the presence of mutation and selection and obtain
the governing equations, which determine the next fractional populations from
the previous ones. The governing equations will be used to investigate the
problems of genetic propagations of chromosome-linked diseases such as color
blindness and hemophilia.

\section{fractional population equations}
We consider the five fractional populations with the following constraints:
\begin{eqnarray} \label{eq:consts}
 XX(n)+X^\prime X(n)+X^\prime X^\prime(n)=1 \\ \nonumber
 XY(n)+X^\prime Y(n)=1,
\end{eqnarray}
for the $n$-th generation. The ratios of gene frequencies without mutation
can be determined as
\begin{eqnarray} \label{eq:genes}
\widetilde{X_f}(n) &=& XX(n) + \frac12 X^\prime X(n), \nonumber \\
\widetilde{X_f^\prime}(n) &=& X^\prime X^\prime(n) + \frac12 X^\prime X(n), \\
\widetilde{X_m}(n) &=& XY(n), \nonumber \\
\widetilde{X_m^\prime}(n) &=& X^\prime Y(n). \nonumber
\end{eqnarray}

Although the mutation between the normal chromosome $X$ and the defective one $X^\prime$
would be rare, still it plays an important role in the following discussion.
In order to consider the mutation, we introduce two probability factors,
$\alpha$ and $\beta$ for the following mutation processes;
\begin{equation}
X\stackrel{\alpha}{\longrightarrow}X^\prime \qquad \mbox{and} \qquad
X\stackrel{\beta}{\longleftarrow}X^\prime.
\end{equation}
Through the above mutation processes, the gene frequencies are modified as
\begin{eqnarray} \label{eq:final}
 X_f(n) &=& (1-\alpha)\widetilde{X}_f(n) +
 \beta\widetilde{X}_f^\prime(n), \nonumber \\
 X_f^\prime(n) &=& (1-\beta)\widetilde{X}_f^\prime(n) +
 \alpha\widetilde{X}_f(n), \\
 X_m(n) &=& (1-\alpha)\widetilde{X}_m(n) +
 \beta\widetilde{X}_m^\prime(n), \nonumber \\
 X_m^\prime(n) &=& (1-\beta)\widetilde{X}_m^\prime(n) +
 \alpha\widetilde{X}_m(n). \nonumber
\end{eqnarray}

The fractional population equations, which govern the populations of the next
generation, are now written as
\begin{eqnarray} \label{eq:next}
\widetilde{XX}(n+1) &=& X_{f}(n) \times X_{m}(n), \nonumber \\
\widetilde{X^\prime X}(n+1) &=& X_{f}^\prime (n)\times X_{m}(n)
                             +X_{f}(n)\times X_{m}^\prime (n), \nonumber \\
\widetilde{X^\prime X^\prime} (n+1) &=& X_{f}^\prime (n)\times
                                X_{m}^\prime (n), \\ \nonumber
\widetilde{XY}(n+1) &=& X_{f}(n), \\ \nonumber
\widetilde{X^\prime Y}(n+1) &=& X_{f}^\prime (n).
\end{eqnarray}

Since the defective chromosome causes a disease, the populations of
$\widetilde{X^\prime X^\prime}$ and $\widetilde{X^\prime Y}$ will have less
chances of surviving and inheriting their genes. In order to reflect this
disadvantages, we introduce {\it disadvantage factors} $\delta_{f}$ for the
female and $\delta_{m}$ for the male groups, respectively. Then, the
populations of $\widetilde{X^\prime X^\prime}$ and $\widetilde{X^\prime Y}$ will
be modified as
\begin{eqnarray}
\mbox{Pop}[\widetilde{X^\prime X^\prime}(n+1)]
&\longrightarrow& (1-\delta_f)\mbox{Pop}[\widetilde{X^\prime X^\prime}(n+1)],\\
\mbox{Pop}[\widetilde{X^\prime Y}(n+1)]
&\longrightarrow& (1-\delta_m)\mbox{Pop}[\widetilde{X^\prime Y}(n+1)]. \nonumber
\end{eqnarray}
With normalization, the fractional population equations are given by
\begin{eqnarray} \label{eq:modify}
XX(n+1) &=& \frac{X_f(n) \cdot X_m(n)}
                    {1-\delta_f X_f^\prime(n)\cdot X_m^\prime(n)}, \nonumber \\
X^\prime X(n+1) &=& \frac{X_f^\prime(n)\cdot X_m(n)
                             +X_f(n)\cdot X_m^\prime(n)}
                 {1-\delta_f X_f^\prime(n)\cdot X_m^\prime(n)}, \nonumber \\
X^\prime X^\prime(n+1) &=& \frac{(1-\delta_f)X_f^\prime(n)\cdot
                                X_m^\prime(n)}
                 {1-\delta_f X_f^\prime(n)\cdot X_m^\prime(n)}, \\ \nonumber
XY(n+1) &=& \frac{X_f(n)}{1-\delta_m X_f^\prime(n)}, \\ \nonumber
X^\prime Y(n+1) &=& \frac{(1-\delta_m)X_f^\prime(n)}
                 {1-\delta_m X_f^\prime(n)}.
\end{eqnarray}
The above governing equations (\ref{eq:consts})$\sim$(\ref{eq:modify}) yield the
following constraint relations for any generation $n$,
\begin{equation}
X_{f}(n)+X_{f}^\prime (n)=1 \qquad \mbox{and} \qquad
X_{m}(n)+X_{m}^\prime (n)=1.
\end{equation}

In order to understand the change of populations along generations, it is
convenient to consider the automata equations for $X_f^\prime$ and $X_m^\prime$
only, which are given by
\begin{eqnarray} \label{eq:auto}
X_m^\prime(n+1) &=&
   \frac{\alpha + \left(1-\alpha-\beta-\delta_m(1-\beta)\right)X_f^\prime(n)}
        {1-\delta_m X_f^\prime(n)}, \\ \nonumber
2X_f^\prime(n+1) &=&
   \frac{2\alpha+(1-\alpha-\beta)\left(X_f^\prime(n)+X_m^\prime(n)\right)
          - 2\delta_f(1-\beta)X_f^\prime(n)\cdot X_m^\prime(n)}
        {1-\delta_fX_f^\prime(n)\cdot X_m^\prime(n)}.
\end{eqnarray}
The above coupled recursion relations can now be used to study the fixed points
of $X_m^\prime$ and $X_f^\prime$, which correspond to the equilibrium values
where $X_m^\prime(n+1)=X_m^\prime(n)={X_m^\prime}^\ast$ and
$X_f^\prime(n+1)=X_f^\prime(n)={X_f^\prime}^\ast$. The fixed points of
${X_f^\prime}^\ast$ are given by the solutions of the algebraic equation,
\begin{equation} \label{eq:general}
 a_0 + a_1 {X_f^\prime}^\ast + a_2 ({X_f^\prime}^\ast)^2
+ a_3 ({X_f^\prime}^\ast)^3 = 0,
\end{equation}
where the coefficients $a_{i}$ are given by
\begin{eqnarray} \label{eq:coeffs}
 a_0 &=& \alpha(3-\alpha-\beta), \nonumber \\
 a_1 &=& (\alpha+\beta)(\alpha+\beta-3)+\delta_m(2\beta-\alpha-\beta^2
         -\alpha\beta-1) +2\delta_f\alpha(\beta-1), \\ \nonumber
 a_2 &=& \delta_m(1+\alpha+\beta)+2\delta_f(2\alpha+2\beta-\beta^2-\alpha\beta
       -1)+2\delta_m\delta_f(\beta-1)^2,  \\ \nonumber
 a_3 &=& 2\delta_f(1-\alpha-\beta)+2\delta_m\delta_f(\beta-1). \nonumber
\end{eqnarray}
Solving this equation for stable fixed points, we can readily determine the
equilibrium population ratios.

We study the fixed points in several cases. First of all, in the case of no
mutation; $\alpha=0$ and $\beta=0$, the meaningful fixed point,
${X_f^\prime}^\ast$, is given by 0. It correctly predicts that without mutations,
the defective genes will disappear eventually. Secondly, when only mutations are
considered in the theory assuming $\delta_m = 0$ and $\delta_f = 0$, the fixed
point is given by ${X_f^\prime}^\ast=\alpha / (\alpha + \beta)$.  This result will
be used in the discussion for color blindness. In other general cases, the exact
solution cannot be expressed in a closed form. However, since the mutation rates
are known to be very small($\sim 10^{-5} \sim 10^{-7}$), the fixed point
${X_f^\prime}^\ast$ can be expressed in terms of $\alpha$ and $\beta$ in an
approximate fashion and will be discussed in the next section.

\section{color blindness and hemophilia}
The disadvantage, that a color blindness man or woman has, is not severe enough to
reduce the chance of survival significantly. Hence, we let the disadvantage factors
be simply zero in this case. Then, we easily notice from
Eq. (\ref{eq:auto}) that the fixed point is given by
\begin{equation}
{X_m^\prime}^\ast={X_f^\prime}^\ast=\frac{\alpha}{\alpha + \beta}.
\end{equation}
This result is identical to that obtained in the conventional genetics\cite{hedrick}.

Using the fractional population equations of Eq. (\ref{eq:modify})
and the above fixed point, we obtain the population ratios as
\begin{equation}
XY:X^\prime Y = X_f^\ast:{X_f^\prime}^\ast =
 1-{X_f^\prime}^\ast:{X_f^\prime}^\ast
=\beta:\alpha,
\end{equation}
and furthermore
\begin{equation}\label{eq:HWlaw}
XX:X^\prime X:X^\prime X^\prime =\beta^{2}:2\beta\alpha:\alpha^{2}.
\end{equation}
The above result is the well known the Hardy-Weinberg law\cite{Hardy-Weinberg}.

The demographic data for color blindness in England show that
$XY:X^\prime Y=12:1$\cite{color}. Hence we conclude that $\beta:\alpha=12:1$.
We notice that $X^\prime $ chromosome is more unstable than $X$ chromosome since $\beta$
is much larger than $\alpha$. The abundance of carrier female is easily noticed by
the fact that $XX:X^\prime X:X^\prime X^\prime =144:24:1$.

Hemophilia is a dreadful disease which affects the chance of survival and
mating significantly. All of the female group $X^\prime X^\prime$ perish completely upon
birth. Hence, the disadvantage factor $\delta_f$ is equal to 1. Then the fixed
point ${X_f^\prime}^\ast$ can be expressed up to the second order of $\alpha$
and $\beta$ as follows using Eq. (\ref{eq:general}) and (\ref{eq:coeffs}),
\begin{eqnarray} \label{eq:hemo}
{X_f^\prime}^\ast &=& \frac3\delta_m \alpha +
  \left(-\frac4\delta_m + \frac{12}{\delta_m^2} -
         \frac{18}{\delta_m^3} \right) \alpha^2 +
  \left( \frac5\delta_m - \frac9{\delta_m^2} \right) \alpha\beta +
  {\cal O}(\alpha^3, \alpha^2\beta, \alpha\beta^2, \beta^3), \\
\label{eq:hemo_m}
{X_m^\prime}^\ast &=& (-2+\frac3\delta_m )\alpha +
  \left(-2-\frac{10}{\delta_m} + \frac{30}{\delta_m^2} -
         \frac{18}{\delta_m^3} \right) \alpha^2 +
  \left(-2+ \frac{11}{\delta_m} - \frac{9}{\delta_m^2} \right) \alpha\beta +
  {\cal O}(\alpha^3, \cdots).
\end{eqnarray}
A numerical calculation of the stable fixed point, ${X_f^\prime}^\ast$ using
Eq. (\ref{eq:general}) is shown in Fig. 1. We find that the dominant
contributions come from $\alpha$ and $\delta_m$ as Eq. (\ref{eq:hemo})
indicates. We also find that the values of the fixed point ${X_f^\prime}^\ast$
are independent of initial values. The approximate expressions, Eqs.
(\ref{eq:hemo}) and (\ref{eq:hemo_m}) are found in good agreement with the
exact expression Eq. (\ref{eq:general}) except when $\delta_m$ is near zero.

It is useful to consider the male hemophilia population before selection in order
to relate the above formulation with the statistical data. Here, the relevant
demographic data is the ratio, $r$, between the mutation cases and the all
hemophilia cases; $\widetilde{X^{\prime}Y}^{\ast}_{mut}
=r\cdot \widetilde{X^{\prime}Y}^{\ast}$. Using Eqs.
(\ref{eq:genes})$\sim$(\ref{eq:modify}), the population of male infants with
hemophilia before selection can be written as
\begin{equation}
\widetilde{X^{\prime}Y}^{\ast}=\frac{1}{2}(1+\alpha -\beta)
X^\prime X^{\ast}+\alpha XX^{\ast}\equiv
\widetilde{X^{\prime}Y}^{\ast}_{inh}+\widetilde{X^{\prime}Y}^{\ast}_{mut}.
\end{equation}
Here, the second term represents the male population having hemophilia caused by
the spontaneous gene mutation. Actually, the first term also contains the gene
mutation contribution of $\frac12(\alpha-\beta)X^\prime X^{\ast}$. However,
when the demographic data $r$ are collected, there is no way to distinguish the
inheritance from the mutation in this case, because data collectors simply check whether
there were hemophiliac occurrences in the family line or not. It is straightforward
to show that  $r$ is related to the disadvantage factor $\delta_{m}$ and the
mutation rates $\alpha$ and $\beta$. From Eqs.
(\ref{eq:genes})$\sim$(\ref{eq:modify}), we find
\begin{equation}\label{eq:r}
\delta_m = 3r + (10-10r-\frac2r)\alpha + (-3+5r)\beta + {\cal O}(\alpha^2,
\alpha\beta).
\end{equation}

The early statistical data for the infant male population having
hemophilia\cite{hemo} show that
$\widetilde{XY}:\widetilde{X^\prime Y}\simeq10^{4}:1$. Assuming that the current
fractional population distribution has reached a fixed point,
we find ${X_f^\prime}^\ast\simeq10^{-4}$ from Eq. (\ref{eq:next}).
Furthermore, a recent statistics shows that the rate $r$ is given by
$r\simeq\frac13$\cite{web}. This data and Eq. (\ref{eq:r}) yield
$\delta_{m}=1-{\cal O}(\alpha,\beta)$. This result is in a reasonable agreement
with the fact that various therapies treating male hemophilia have been
invented only recently and, thus, the probability of successful marriage and
reproduction was almost zero for male patient in the past. With $\delta_m\simeq1$
and ${X_f^\prime}^\ast\simeq10^{-4}$, we find from Eq. (\ref{eq:hemo}) that the
value of $\alpha$ is about $3.3\times 10^{-5}$, which is in a reasonable range as
the probability of mutation. Also, we can obtain the population ratio of the
carrier female group using Eq. (\ref{eq:modify});
$XX:X^\prime X:X^\prime X^\prime \sim 10^{4}:1.33:0$.

Since the level of therapies treating male hemophilia has now reached the stage
that most of male patient can marry and reproduce, it is interesting to study
the case $\delta_m=0$ and make predictions how the mutation rate and the
fractional population will change accordingly. The fixed point of Eq.
(\ref{eq:hemo}) and (\ref{eq:hemo_m}) will be modified as
\begin{eqnarray} \label{eq:hemoall}
{X_f^\prime}^\ast &=& \sqrt{\frac32}\alpha^\frac12 +{\cal O}(\alpha,\cdots), \\
{X_m^\prime}^\ast &=& \alpha + (1 - \alpha + \beta ){X_f^\prime}^{\ast}.
\end{eqnarray}
Using the above results, we find the modified rate
\begin{equation}
r = \sqrt{\frac23}\alpha^\frac12 + {\cal O}(\alpha,\cdots).
\end{equation}
Assuming the mutation rate $\alpha$ does not change and remain
$\alpha \simeq 3.3\times 10^{-5}$ as we find in the above, we can
determine ${X_f^\prime}^\ast=0.00707$. Also, we readily obtain the fractional populations:
$XX:X^\prime X:X^\prime X^\prime =98.6:1.41:0$, and $XY:X^\prime Y=993:7$. The
result clearly shows that the majority of hemophilia would results from
inheritance and that the fractional population of carrier female increases
drastically, when male hemophiliac patients survive and mate without any
disadvantages. Therefore, nongenetic treatment of hemophiliac male may cause
increase of hemophiliac population and infant deaths of female patients, unless
some concurrent measures are taken. However, we note that the present
calculation can not produce the dynamic properties of the transition period
between $\delta_m\simeq1$ and $\delta_m\simeq0$, since $\delta_m$ is assumed
static in the calculation.

\section{conclusion}
We have considered the population ratios of the genetic groups related with
chromosome inherited diseases. The governing equations, which determine the
ratios, are found in the presence of mutation and selection. The selection is
taken into account in the formulation by using the disadvantage factors. It is
found that there exist physical fixed points in the automata equations, which
correspond to equilibrium population rates. These fractional population
equations are used to discuss the cases of color blindness and hemophilia.

In the case of color blindness, there is no significant disadvantage in
selection so that the disadvantage factors can be assumed zero. From the governing
equations, we readily obtain the Hardy-Weinberg relation
$XX:X^\prime X:X^\prime X^\prime=\beta^{2}:2\alpha\beta:\alpha^{2}$. Using the
statistical data $XY:X^\prime Y=12:1$, we find that the ratio of mutation
rates $\alpha:\beta=1:12$.

Hemophilia seriously hampers chances of survival, mating and reproduction.
Especially for female patients, chance of survival is almost zero, thus, making
$\delta_f=1$. Using the demographic data that one out of ten thousand  male
infants have hemophilia, and that one third of all hemophiliac cases are thought
to be caused by gene mutation, we obtain the following results;
i) the disadvantage factor $\delta_{m}$ for male is almost 1,
ii) the mutation rate $\alpha\simeq3.3\times 10^{-5}$, and
iii) $XX : X^\prime X \simeq 10^4 : 1.33$.

We have also studied the case when the hemophiliac male suffers no disadvantage in
the selection process; $\delta_m=0$. It is found that the population of
hemophiliac females and males would increase drastically and inheritary
hemophilia would be dominant over gene mutated cases.

\vspace{1cm}
\acknowledgments
This work has been partly supported by the Korea Ministry of Education
(Grant No. BSRI-97-2425) and the Korea Science and Engineering Foundation
through Project No. 95-0701-04-01-3 and also through the SRC program of SNU-CTP.

\begin{figure}
\centering{\epsffile{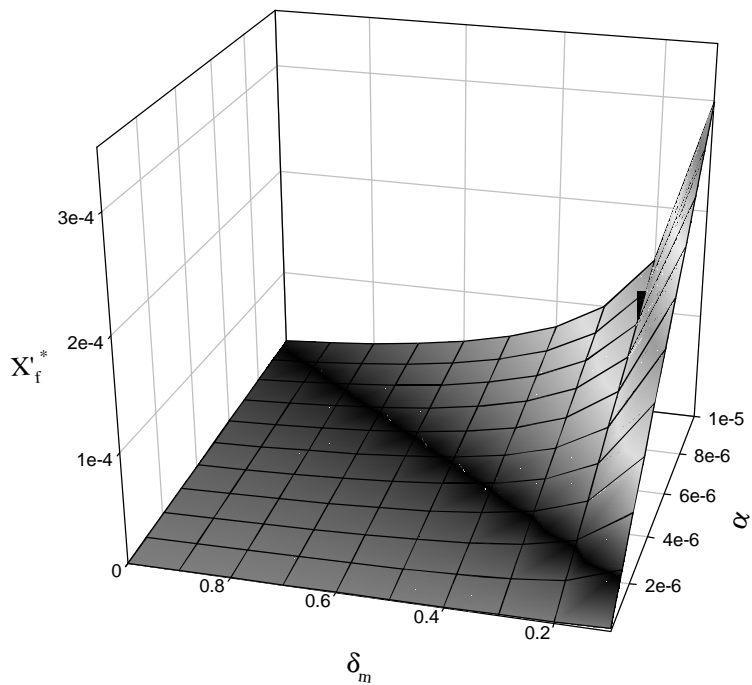}}
\vspace{5mm}

\caption{For a set of values of $\delta_f=0.5$ and $\beta = 10^{-5}$, we plot
the fixed point ${X_f^\prime}^\ast$ in the three dimensional format, where the
$x$ and $y$-axis correspond the mutation rate $\alpha$ and the male disadvantage
factor $\delta_{m}$. It is found that the overall features of the shape and size do
not depend on $\delta_f$ and $\beta$ sensitively.}
\end{figure}

\end{document}